# Evaporation of water in a microfluidic channel under magnetic field.

Sruthy Poulose[1], Yara Alvarez-Braña[2,3], Lourdes Basabel-Desmonts[3,4], Fernando Benito-Lopez[2] and J. M. D. Coey[1]

[1] School of Physics, Trinity College, Dublin 2, Ireland.

[2] Microfluidics Cluster UPV/EHU, Analytical Microsystems & Materials for Lab-on-a-Chip Group, Analytical Chemistry Department, University of the Basque Country UPV/EHU, Spain

[3] Microfluidics Cluster UPV/EHU, BIOMICs microfluidics Group, Lascaray Research Center, University of the Basque Country UPV/EHU, Vitoria-Gasteiz, Spain

[4] IKERBASQUE, Basque Foundation for Science, Bilbao, Spain

**Abstract.**

The evaporation of drops of water placed at the center of long poly(methyl methacrylate) microfluidic channels with a rectangular cross section of 0.38 mm$^2$ is studied by simultaneously monitoring the shapes of two samples, one is in a 300 mT magnetic field, the other is in no field. A magnetic enhancement of the evaporation rate of up to 140 % is observed, which can be understood by treating the *ortho* and *para* nuclear isomers of water vapor as quasi-independent gasses with an *ortho:para* ratio in fresh vapor close to 2:3. It would take much longer than the 2 - 4 h duration of an experiment in the channel, for the ratio to approach the 3:1 equilibrium value. Magnetic field influences evaporation rate by equalizing the isomeric populations in the vapor phase. The atmosphere in the channel is saturated with water vapor yet the evaporation rate far exceeds that in open beakers.

jcoey@tcd.ie

Microfluidics uses microliter sample volumes in channels with sub-square-millimeter cross sections for a wide variety of applications [1]. The fluids used are often water-based, and the behavior of water in these small channels is a continuing topic of discussion, where forces that scale with the first or second powers of the dimensions, such as surface tension γ and viscosity μ, dominate body forces such as gravity ρg, and inertia. The capillary length below which body forces are no longer dominant is $l_c = \sqrt{(\gamma/\rho g)}$, where ρ is the density, The value of $l_c$ for water is 2.7 mm.

Loss by evaporation is usually negligible in a miniature channel on the timescale of a microfluidic measurement, but evaporation-induced flow is prominent in microscale environments such as ldroplets, menisci, thin film and liquid or solid reagent storage. The evaporation and interfacial flow of ethanol or methanol at a meniscus in a horizontal capillary tube has been studied by particle image velocimetry (⌈PIV) [2 - 4]. A circulating flow with two counter-rotating vortices is observed near the evaporating meniscus in a horizontal plane, whereas in a vertical plane where gravity does exert an influence, a single vortex is seen. Non-uniform evaporation across a meniscus, fastest at the triple line where vapor, liquid and solid meet [4,5], leads to nonuniform cooling and an evaporation-induced temperature gradient, which in turn creates a surface tension gradient along the liquid-air interface that leads to vortex flow — the thermocapillary Marangoni effect. Thermocapillarity is exploited in microfluidic devices [6]

Much effort has been devoted to studying the evaporation of microliter drops of water, which is of interest in areas ranging from atmospheric physics to digital microfluidics and spray coating. Studies by Ward and Duan presented evidence of direct thermocapillary flow in water with a surface velocity of order 1 mms$^{-1}$ to replenish the loss by evaporation at the triple line. They used a small metallic funnel [7,8], trough [9] or flat substrate [10] and eliminated gravitational convection by controlling the temperature to just below 4 °C. The Marangoni number Ma = -(∂γ/∂T)LΔT/μα was above 100; ΔT/L is the temperature gradient across the surface, and α is the thermal diffusivity of water. Evaporation, especially at the triple line, creates surface temperature gradients and corresponding gradients of surface tension that might be expected to induce Marangoni convection within water, like that observed in alcohol. Local thermal gradients created in a sessile droplet [11,12,13], for example by heating the center by a degree or two with laser light passing through the substrate [11], do initiate or maintain Bénard-Marangoni vortex flow at velocities of 30 mms$^{-1}$ both in the laboratory and in simulations [14,15], which emphasize the importance of taking airflow in the ambient atmosphere into account. It is possible to obtain transient effects without external heating in a spherical sessile or pendant droplet on a thin copper rod [16]. The flow may be detected by particle image velocimetry, but results are very sensitive to traces of surfactant or impurities in the water and the effect is not always seen [15, 17]

Various claims have been made for effects of a magnetic field on the physical properties of water [18]. An enhancement of the evaporation rate in a static field has been measured in different ways [18-21] but its molecular origin was perplexing. A recent study, where we monitored the evaporation simultaneously from two beakers of water kept in the same environment, one in a 500 mT field and the other in no field, established a $12 \pm 7$ % increase of evaporation rate in the presence of the field [22]. By analogy with *ortho* and *para* hydrogen, where the total nuclear spin of the two protons in H$_2$ is I = 1 or 0, respectively [23], the field effect in H$_2$O was explained by treating the *ortho* and *para* nuclear isomers of water vapor as quasi-independent gasses. The ortho fraction in freshly evaporated water vapor was found to be $39 \pm 1$ % [22], compared to the value of 75 % in ambient air — the equilibrium 3:1 *ortho:para* ratio. The magnetic field changes the evaporation rate by altering the isomeric ratio in the vapor. Here we apply these ideas to a much stronger influence of magnetic field on the evaporation of water that is confined in a microfluidic channel.

Our channels are made from three layers of hydrophobic poly(methyl methacrylate) (PMMA). The microfluidic chips are assembled by thermo-lamination after cutting a 1 mm wide channel, 54 mm long in the 0.38 mm thick middle layer with a CO$_2$ laser. A schematic image of the experimental setup is shown in Fig. 1. Two chips are used in each experiment. No magnetic field is applied to one of them, while a 300 mT magnetic field is applied to the other using rectangular 50 x 20 x 10 mm$^3$ Nd-Fe-B permanent magnets with a remanence of 1.26 T. They are magnetized along the short axis and used to create a field perpendicular to channel. The evaporation of water in the two chips is monitored simultaneously using two PCE800mm USB cameras. Channels are pre-wetted prior to the measurements by introducing 0.6 – 0.7 microliters of Millipore deionized water at one end and drawing it through the channel with a syringe pump connected to the other end. Then a further droplet of about 0.4 microliters of water is introduced into the pre-wetted channel and brought to the center using the pump. The two ends are left open, and the evolution of the shapes of the drops in the two channels is recorded as they shrink down to a membrane after several hours, which ruptures shortly afterwards. The whole setup was placed inside a perspex box, and temperature (26 °C) and relative humidity RH (42 %) in the laboratory were maintained throughout the experiments.

After moving a 0.4 µL droplet to the center of a channel, two menisci soon form at the edges with contact angles $\vartheta$ of about 30 °, as seen in Fig 2. PMMA is hydrophilic with a contact angle < 90° [24], but the main reason for menisci to form is the periodic variation of the channel width due to cutting with the $CO_2$ laser. The variation is 10 % of the 1 mm width, and the period is 300 µm. Just like a soap film or a magnetic domain wall, the water surface seeks to minimize its area and form a circular meniscus because of surface tension. The contact angle at the laser-cut edges oscillates in the course of the evaporation experiment as a meniscus moves in to the next pinning point in a stick-slip process. The contact angle at the upper and lower PMMA surfaces is close to 90 °. The volume of water remaining in the drop is obtained directly from its approximately symmetric shape, which is parameterized by *L,* the average of contact lengths at the two edges and *l,* the shortest distance between the centers of the two menisci, as shown in Fig. 1c). The volume of liquid is calculated from *L* and *l* knowing the width *w* and the depth *d* of the channel.

The the radius of curvature *R* of the menisci is related to *L*, *l* and *w* by the intersecting chords theorem:

$$R = \tfrac{1}{4}\,[w^2/(L - l) + (L - l)] \qquad (1)$$

The area $A$ of each meniscus, $2Rd(\pi/2 - \vartheta)$ is

$$A \approx 2Rd\,\sin^{-1}(w/2R) \qquad (2)$$

and a good approximation for the volume of the space spanned by the meniscus for a wide range of contact angle $20° < \theta < 60°$ is $w^2 d\theta/3$. Hence

$$V \approx wd\,(L - 2w\vartheta/3) \qquad (3)$$

The contact angle $\theta = \tan^{-1}[(2R-L+l)/w]$. For example, the volume of a sample of water with $L = 2$ mm in the channel of width $w = 1$ mm and depth $d = 0.38$ mm, when $\theta = 0.5^c$ (29 °) is 0.51 µL. R is 0.57 mm and $l = 1.4$ mm. Snapshots of the evolving meniscus are shown in Fig 2. The curved shape of the surfaces helps to increase the evaporation rate, but the meniscus area, where evaporation is occurring, remains roughly constant and equal to $2wd$.

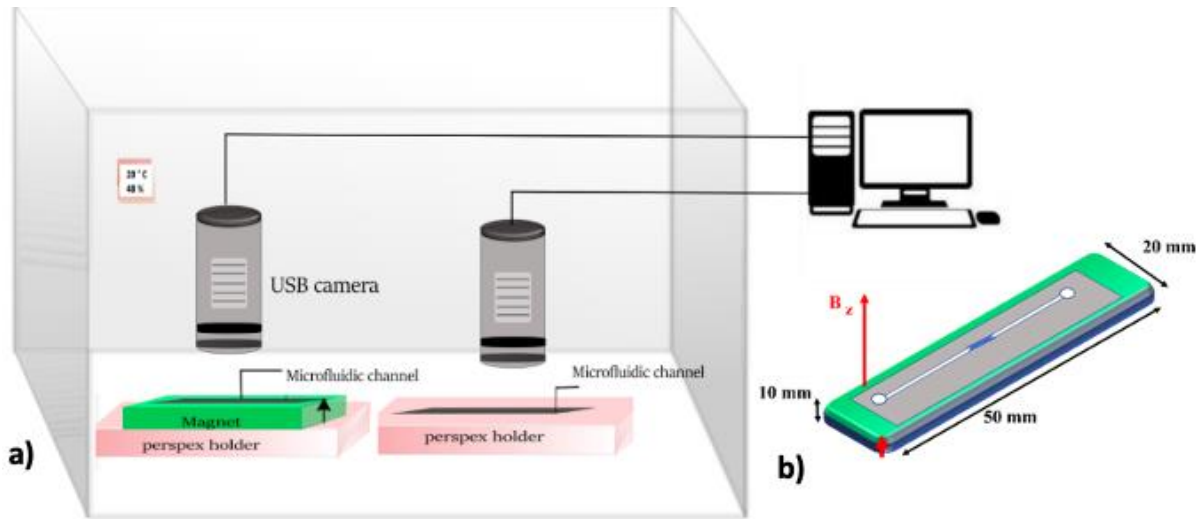

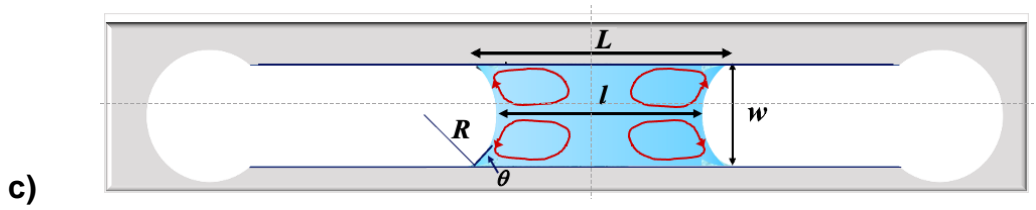

c)

Figure 1: a, b) Schematic of the experimental arrangement, and c) dimensions of a sample of water in the channel (top view). The menisci are caused by periodic variation of the channel width *w*, which is visible in Fig. 2 The proposed Marangoni flow vortices are indicated in red.

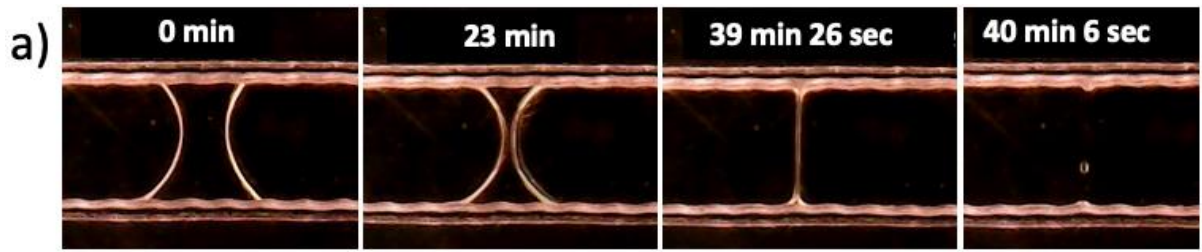

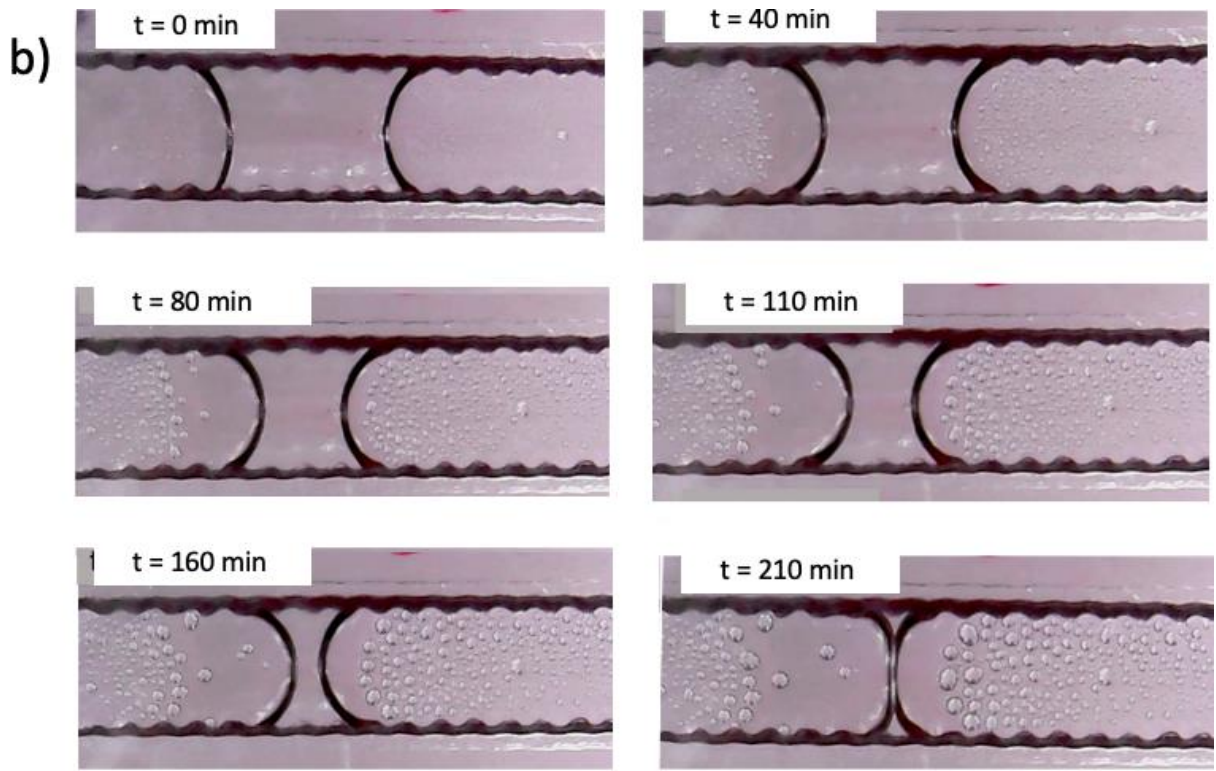

Figure 2: a) Evolution of 0.12 µL drop of water in a microfluidic channel. It shrinks in 39 min to leave a membrane that ruptures 30 s later. b) Another example, of a 0.51µL drop. Note the growing droplets of recondensed water vapor forming in the channel.

Data shown in Fig. 3 are for two representative experiments where the water drop takes about four hours to evaporate without a field. Initial values of $L$ and $l$ are normalized to 1. The in-field evaporation rate is considerably faster. The variation of $L$ shows steps related to discontinuous jumps of the menisci from one pinning point to the next, but the time-dependence of $l$ is smoother.

In most cases we observe stable sessile droplets of water with diameters of 30 - 150 µm growing slowly in the channel throughout a run. They are seen at distances greater than about 100 µm from the menisci, (see Fig. 2b) which means that the air in the microchannel at 299 K must be saturated with water vapor. Sessile droplets in arrays are known to evaporate faster in confined conditions in unsaturated air if they are located at the edge of the array than if they are surrounded by other droplets [25]. Here we observe the converse effect of growth in saturated vapor

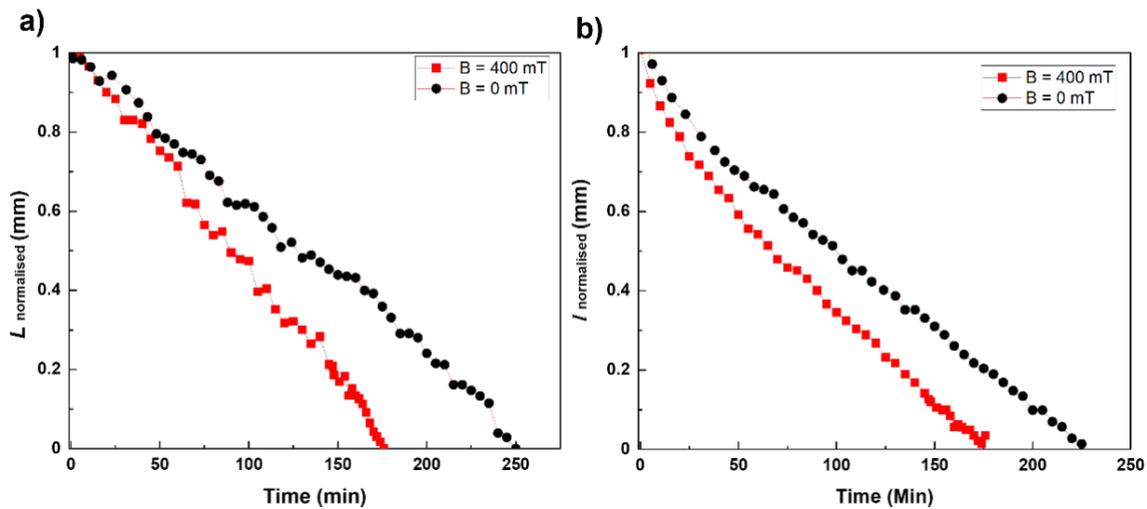

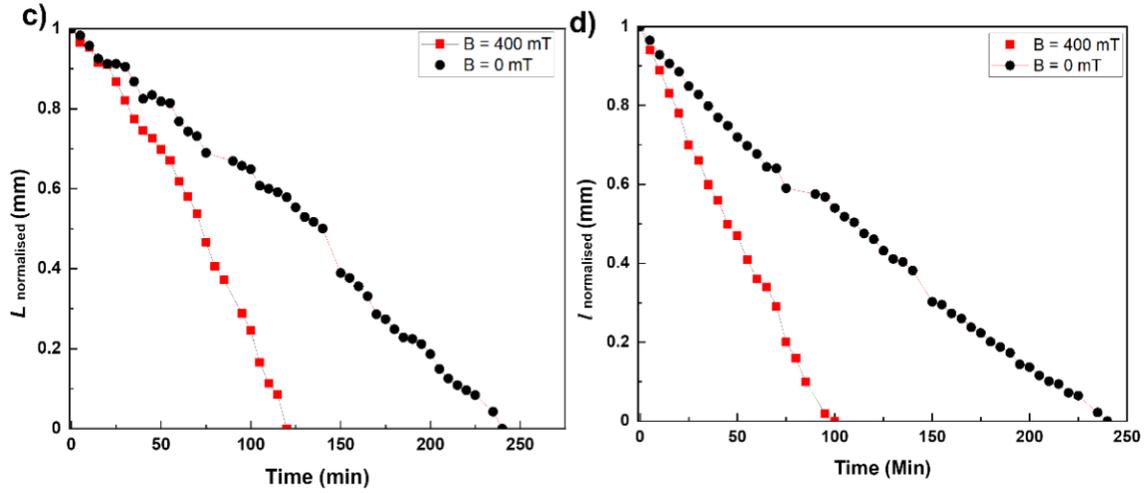

Figure 3: Variation of *L* and *l* during two evaporation experiments; a) and b) are for Experiment, 1 c) and d) are for experiment 2. Data in black are for the no-field channel and data in red are for the channel in a 300 mT magnetic field.

The average evaporation rate for water in the microchannel is approximately 0.132 kgm$^{-2}$ h$^{-1}$, two to four times *faster* than the evaporation rate from half-filled beakers [22] where the water also evaporates into its own vapor, but the air is there is unsaturated, and the rate decreases strongly with ambient relative humidity in the range 50 – 80 % [22]. The flow dynamics of water in the microchannel and in the beaker must be quite different. The flow regime may be characterized by the inverse Bond number, the dimensionless ratio $-(\partial\gamma/\partial T)/\beta\rho g d^2$ of Marangoni to Rayleigh numbers, which reflects the relative importance of surface tension and gravitational forces in influencing the flow. Here, $\beta = 2.1 \times 10^{-4}$ K$^{-1}$ is the thermal expansion coefficient of water and *d* is the depth. The ratio is 316,000 for the microchannels and 90 for the beakers.

The vapor close to the meniscus is composed of freshly-evaporated molecules, which need high kinetic energy and well-timed making and breaking of at least three hydrogen bonds in the interface in order to break free [26]. The effective temperature of the velocity distribution of such water vapor will be higher than ambient, and it will not immediately condense. We see a Knudsen layer that is much wider than usual [27]. There must be a temperature gradient normal to the water surface.

We suggest that the enhanced evaporation rate of the water drop confined at the middle of the microchannel may be attributed to Marangoni convection. Cooling by evaporation is most intense at the meniscus edge, which reduces the temperature there and leads to gradients of surface tension at the liquid surface, increasing from the center to the edges, create the two-dimensional twin vortex flow patterns that are sketched in Fig. 1c). They resemble those that have been seen at a meniscus in a capillary [2 - 5].

Turning now to the magnetic field effect on evaporation, the average evaporation rate from the control channel in ten datasets is $0.14 \pm 0.03$ kgm$^{-2}$s$^{-1}$. In different runs the magnetic field enhancement of the evaporation rate ranged up to 140 %, with an average value of $61 \pm 42$ %. The effect is much greater than the one we found in experiments on water evaporation from half-filled 100 mL beakers, where the largest enhancement of the evaporation rate was 36 %, and there the average over 36 runs was $12 \pm 7$ %. [22]. We developed a model to explain these experiments based on the idea that the *ortho* and *para* isomers of water vapor behave as independent gasses, each with its own partial pressure. In that work, the ortho percentage $f^o{}_L$ in fresh water vapor was found to be $39 \pm 1$ %, compared to the equilibrium value of 75 % in ambient air, which is only approached very slowly because transitions between the two isomeric states are strongly forbidden [23]. The water vapor in the confined space in the microchannel on either side of the water will therefore have the same composition as freshly-evaporated vapor. Treating the two as independent gasses, we propose the ansatz that the evaporation rate will be proportional to a dimensionless sum of the *ortho* and *para* contributions,

$$g_0 = [f^o{}_L(1 - f^o{}_V) + f^p{}_L(1 - f^p{}_V)] \qquad (4)$$

where $f^o$ and $f^p$ are the *ortho* and *para* fractions and L refers to the fresh vapor emanating from the liquid and V refers to the ambient vapor in the channel. For each isomer the evaporation depends on the product of an attempt frequency proportional to $f_L$ and a success rate proportional to (1 - $f_V$), depending on the capacity of the air to absorb the isomer. Since ($f°_L$ + $f^P_L$) = ($f°_V$ + $f^P_V$) = 1, it follows that when the evaporating liquid in the channel is evaporating into its own vapor,

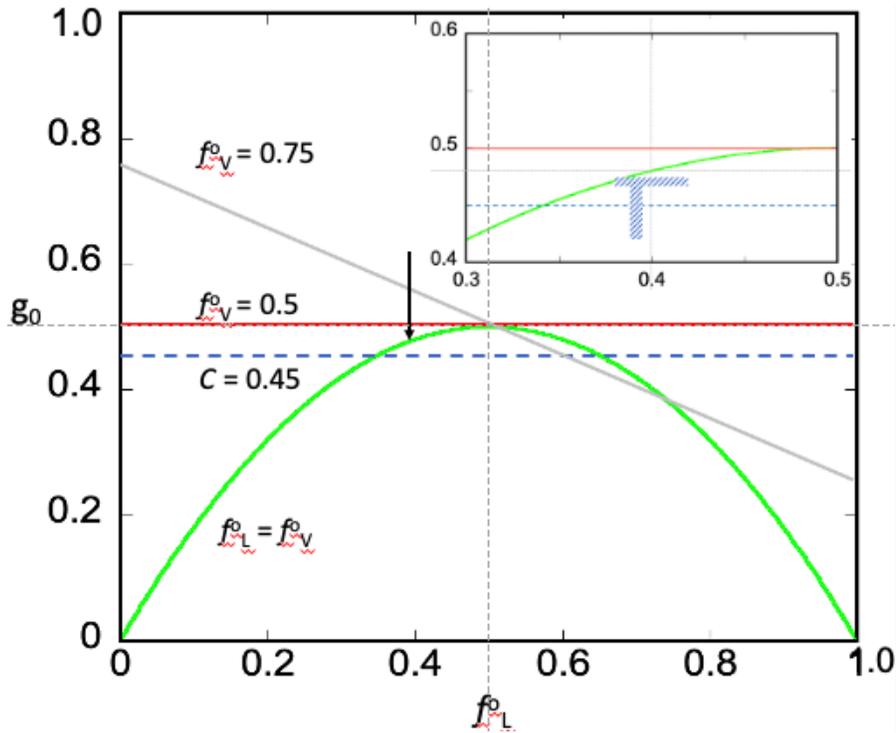

Figure 4: Dimensionless evaporation rates of water as a function of the *ortho* fraction $f°_L$ present in the vapor escaping for the liquid for two cases. The green curve is for $f°_L = f°_V$, when the water is surrounded by its own vapor. The red line is for $f°_V = ½$, after exposure to the magnetic field. The shaded areas indicate the ranges of $f°_L$ and recombination rate $c$ (also read on the vertical axis) that are compatible with the data. The net evaporation rate is proportional to the distance from the dashed blue line to the green curve in zero field or the red line in a magnetic field.

$$g_0 = 2 f°_L (1 - f°_L) \qquad (5)$$

This is the solid green curve plotted in Fig. 4. The effect of Larmor precession of the magnetic moments in a nonuniform magnetic field, or Lorentz torque on the electric dipole moments of the water molecules, tends to equalize $f^o{}_v$ and $f^p{}_v$ [22] so that each approached ½; $f^o{}_L$ = ½ is the horizontal red line in the figure. A key point is that the evaporation rate of pure water always increases in a magnetic field. The measured evaporation is the *net* rate, proportional to $(g_0 - c)$, where $c$ is the relative condensation rate of water vapor re-entering the liquid. Previously [22] we found $f^o{}_L = 39 \pm 1$ % with $c = 0.45$, denoted by the arrow in Fig. 4; the present range of magnetic field effects can be explained in the model if $f^o{}_L = 40 \pm 2$ %. with $c = 0.47$. Otherwise, if the *ortho* isomer content is fixed at 39 % there is a range of recombination rates $0.42 < c < 0.47$.

In conclusion, the average evaporation rate of water in a PMMA microfluidic channel where the relative humidity exceeds 100% is 2 - 4 times greater than in an open beaker. The remarkable enhancement by an applied magnetic field is explained in terms of the *ortho:para* ratio of the evaporated vapor, treating the isomers as independent gasses. Our results suggest that it may be necessary to reconsider the treatment of water vapor as a single gas in stagnant conditions where advection is limited. Magnetic field may prove to be a useful handle on evaporation of water in microscale porous media that is independent of temperature. The strong Marangoni vortex flow inferred in water samples evaporating in a microfluidic channel should be visualized in future work using particle image velocimetry, which could also show whether there is any charge flow associated with the two counter-rotating vortices in water, via the Lorentz force.

**Acknowledgements.** The work was supported by the European Commission from Contract No 766007 for the 'Magnetism and Microfluidics' Marie Curie International Training Network. JMDC acknowledges support from Science Foundation Ireland, contract 12/RC/2278_P2 AMBER. LBD and FBL acknowledge support from the Spanish Ministry of Science and Education under grant PID2020-120313GB-I00 / AIE / 10.13039/501100011033 and by FEDER, and a grant from the Department of Education of the Basque Government for consolidation of the research group (IT1633-22)